\documentclass[a4paper,11pt]{article}
\usepackage{pos}
\usepackage{slashed}
\usepackage{subfig}
\usepackage{booktabs}
\newcommand{\FI}{\mathcal{I}}
\newcommand{\TOP}[1][top]{\mathsf{#1}}
\newcommand{\PL}{\TOP[PL]}

\newcommand{\NA}{\TOP[NA]}
\newcommand{\NB}{\TOP[NB]}

\newcommand{\FIPL}[1]{\FI_{\PL}(#1)}

\newcommand{\FINA}[1]{\FI_{\NA}(#1)}
\newcommand{\FINB}[1]{\FI_{\NB}(#1)}

\newcommand{\HyperInt}{\href{http://bitbucket.org/PanzerErik/hyperint/}{\texttt{\textup{HyperInt}}}}
\newcommand{\td}{\mathrm{d}}
\newcommand{\Li}[1]{\operatorname{Li}_{#1}}
\newcommand{\imineq}[2]{\vcenter{\hbox{\includegraphics[height=#2ex]{#1}}}}

\title{Mixed QCD-electroweak corrections to Higgs plus jet production at the LHC}
\ShortTitle{QCD-EW $PP \to H + j$}

\author*[a]{Marco Bonetti}

\affiliation[a]{Institute for Theoretical Particle Physics and Cosmology, RWTH Aachen University, \\Sommerfeldstrasse 16, D-52056 Aachen, Germany}

\emailAdd{bonetti@physik.rwth-aachen.de}

\abstract{The detailed study of the Higgs boson is one of the main tasks of contemporary particle physics. Gluon fusion, the main production channel of Higgs bosons at the LHC, has been successfully modelled in QCD up to $\text{N}^3\text{LO}$. To fully exploit this unprecedented theoretical effort, sub-leading contributions, such as electroweak corrections, must be investigated. I will present the analytic calculations of the gluon- and quark-induced Higgs plus jet amplitudes in mixed QCD-electroweak corrections mediated by light quarks up to order $v \alpha^2 \alpha_S^{3/2}$.

~

TTK-22-24,~P3H-22-074}

\FullConference{%
  Loops and Legs in Quantum Field Theory - LL2022,\\
  25-30 April, 2022\\
  Ettal, Germany
}

\begin{document}
\maketitle

\section{Introduction \& motivations}
\label{sec:intro}

Since its discovery in 2012~\cite{Aad:2012tfa,Chatrchyan:2012ufa}, the Higgs boson has been one of the most studied objects of contemporary particle physics. In order to probe its properties, both unprecedented precision and accuracy in experimental measurements and state-of-the-art theoretical predictions are required.

The LHC is at present the main (and only) Higgs boson factory, where Higgs bosons are produced by colliding protons. In particular, gluon fusion represents the main production mode, contributing alone more than $80\%$ to the total cross section \cite{ParticleDataGroup:2020ssz}. In light of the next High-Luminosity phase of the LHC we expect the experimental relative uncertainty for many gluon-initiated Higgs production modes to decrease to the level of the percent. In order to match this advancement in experimental results, theoretical prediction must provide results with competitive theoretical uncertainties of the order of $\lesssim1\%$.

A huge computational effort has been invested in the last decade on the theoretical description of Higgs boson production through gluon fusion, which is now known up to $\text{N}^3\text{LO}$ in pure QCD in the limit of infinite top-quark mass \cite{Anastasiou:2015vya,Mistlberger:2018etf}. To make full use of such astonishing result, sub-leading contributions have to be properly addressed and included in theoretical predictions.

As explained in \cite{Anastasiou:2016cez,Mistlberger:2018etf}, the remaining sources of uncertainties are given by the lack of $\text{N}^3\text{LO}$ PDF sets, the approximate estimate of NLO mixed QCD-Electroweak corrections, the missing quark-mass effects beyond NLO, and the error associated to the truncation in $1/m_t$ at NNLO. While the $1/m_t$ uncertainty has been recently removed thanks to \cite{Czakon:2020vql}, the remaining uncertainties amount each to $~1\%$.

\begin{figure}
\centering
	\includegraphics[width=0.20\textwidth]{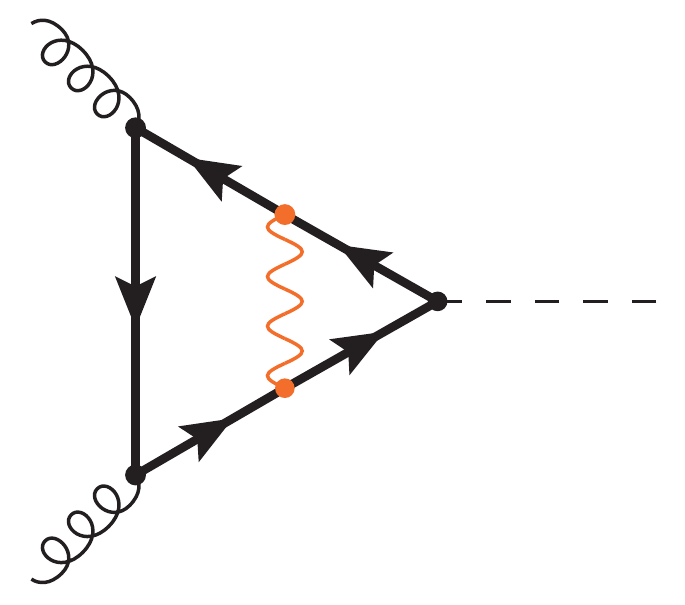}
	\qquad\qquad
	\includegraphics[width=0.20\textwidth]{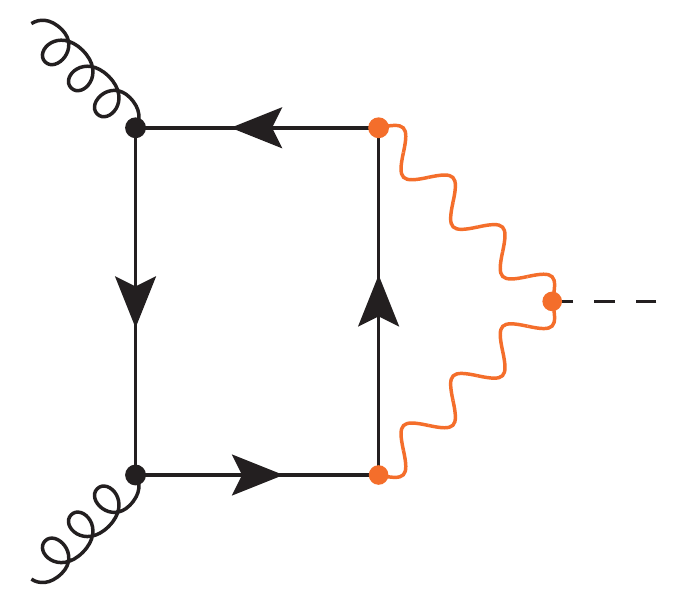}
	\caption{Different LO Electroweak corrections to Higgs boson production in gluon fusion.}
    \label{fig:QCD-EW_modes}
\end{figure}

In this contribution, based on \cite{Bonetti:2020hqh,Bonetti:2022lrk}, we focus on the mixed QCD-Electroweak contributions to Higgs plus jet production at the LHC. Electroweak corrections can contribute with two class of diagrams to gluon fusion, as depicted in Figure~\ref{fig:QCD-EW_modes} \cite{Aglietti:2004nj,Degrassi:2004mx,Aglietti:2006yd}: on the one hand electroweak gauge bosons appear as insertions on the top quark loop connecting the gluons to the Higgs boson, and on the other hand the Electroweak bosons can act as a bridge between the fermion loop and the Higgs boson. At LO, while the first class of contributions, proportional to the top-quark Yukawa coupling, contributes for just half a percent compared to the LO term in pure QCD, the diagrams proportional to $\alpha^2 v$ increase the total cross section by more than $5\%$ of the LO QCD predictions.

Different approximations have been employed to address the NLO mixed QCD-Electroweak corrections, such as Higgs effective field theories employing the limit $m_t,m_W,m_Z \gg m_H$ \cite{Anastasiou:2008tj} or the limit $m_W=m_Z=0$ \cite{Anastasiou:2018adr}. Since QCD effects in Higgs physics can be quite large (up to $+100\%$, as seen in the pure QCD case) such approximations pose a problem both in terms of theoretical uncertainties (from which the $1\%$ mentioned above) and in terms of physical parameters. An analytic computation with full Electroweak boson mass dependence is therefore mandatory to properly control the theoretical uncertainty associated to such contributions.

\section{Tensor structure decomposition}
\label{sec:ampli}

We are interested in the inclusive production of a Higgs boson starting from protons, therefore we need to consider both gluon-initiated and quark-initiated processes in the partonic interaction. The 2-loop LO and 3-loop virtual NLO corrections to $gg \to H$ have been computed in \cite{Aglietti:2004nj,Degrassi:2004mx,Aglietti:2006yd,Bonetti:2016brm} and \cite{Bonetti:2017ovy}, respectively, while the 1-loop LO quark-initiated contributions $q\overline{q}gH$ and the related 1-loop real contributions $q\overline{q}Hgg$ and $q\overline{q}Hq\overline{q}$ have been addressed in \cite{Hirschi:2019fkz}. This leaves the two-loop $ggHg$ and $q\overline{q}Hg$ contributions still to be evaluated.

\begin{figure}
\centering
	\includegraphics[width=0.25\textwidth]{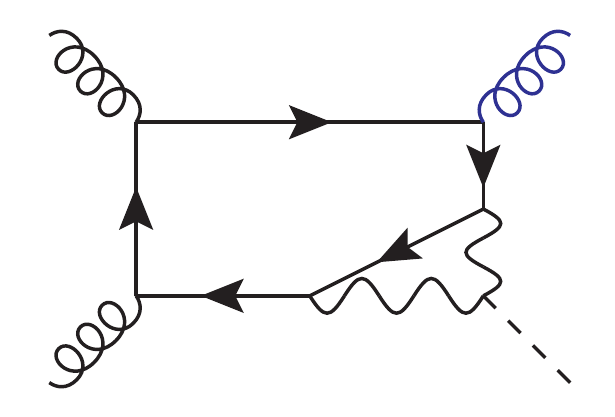}
	\qquad\qquad
	\includegraphics[width=0.25\textwidth]{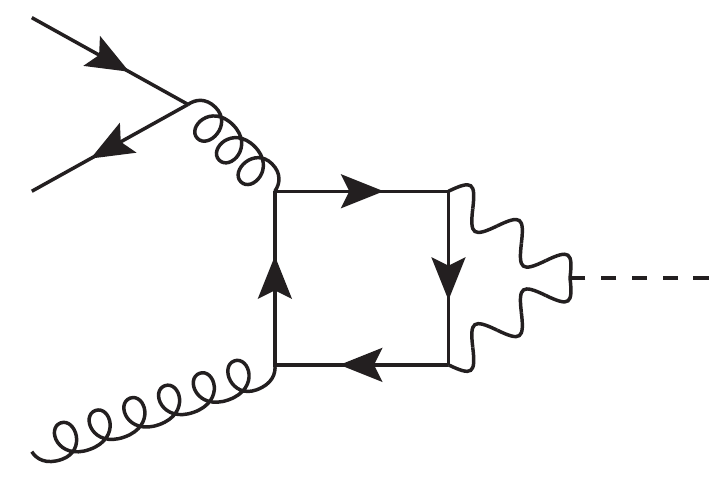}
	\qquad\qquad
	\includegraphics[width=0.25\textwidth]{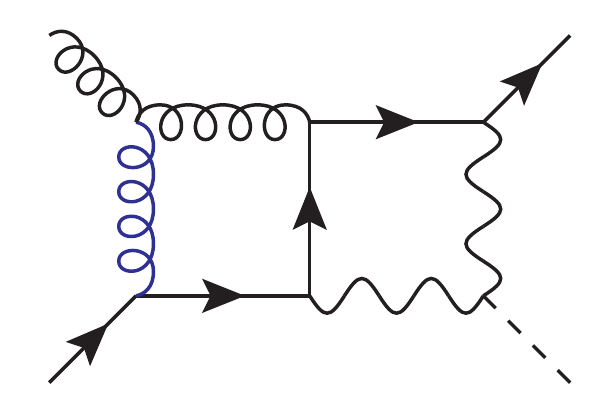}
	\caption{Representative Feynman diagrams for $ggHg$ (left) and $q\overline{q}Hg$ (central and right) contributions. The leftmost and central diagrams feature closed fermion loops, while the rightmost has a single open fermion line.}
    \label{fig:diags}
\end{figure}

We start by decomposing the amplitudes into a linear combination of tensor structures, the coefficients being the form factors. Due to the presence of $W$ and $Z$ bosons interacting with quarks the amplitude might contain terms explicitly depending on $\gamma_5$ or the Levi-Civita pseudo-tensor. When the Electroweak bosons are attached to a closed fermion loop (as in the left and central diagrams of Figure~\ref{fig:diags}) the contributions proportional to $\gamma_5$ cancel themselves once we sum over complete generations of quarks, while in case of a single open fermion line (as in the right diagram of Figure~\ref{fig:diags}) we can consider polarized external states and employ a $\gamma_5$ scheme preserving anti-commutativity to move the $\gamma_5$ until it gets contracted with a polarized spinor. The neat effect of these procedures is that we retrieve the same tensor structures that we would employ in the pure QCD case for the same external states, while the chiral effects are embedded into a rescaling of the coupling constants. We obtain
\begin{align}
    \mathcal{M} &= f^{c_1c_2c_3} \epsilon_{\lambda_1}^\mu(\mathbf{p}_1) \epsilon_{\lambda_2}^\nu(\mathbf{p}_2) \epsilon_{\lambda_3}^\rho(\mathbf{p}_3) \left[ \mathcal{F}_1 g_{\mu\nu}p_{2\rho} + \mathcal{F}_2 g_{\mu\rho} p_{1\nu} + \mathcal{F}_3 g_{\nu\rho} p_{3\mu} + \mathcal{F}_4 p_{3\mu} p_{1\nu} p_{2\rho} \right] ,
\end{align}
\begin{align}
    \mathcal{F}_{1\dots4} &= 4 F_{1\dots4,m_W} + \frac{2}{\cos^4 \theta_W}\left( \frac{5}{4} - \frac{7}{3}\sin^2\theta_W + \frac{22}{9}\sin^4\theta_W \right) F_{1\dots4,m_Z} \,,
\end{align}
for the purely gluonic case and
\begin{align}
    \mathcal{M}^{\text{open}}_L &= T^{c_3}_{i_1i_2} \left( \frac{2}{\cos^4 \theta_W} Q_q^2 \sin^4 \theta_W \right) \overline{v}_{s_1}(\mathbf{p}_1) \mathbb{P}_L \left[ \tau_{1,\mu} A_{1,m_Z}^{\textup{open}} + \tau_{2,\mu} A_{2,m_Z}^{\textup{open}} \right] u_{s_2}(\mathbf{p}_2) \epsilon^{\lambda_3}_\mu(\mathbf{p}_3)  \,,
\end{align}
\begin{align}
    \begin{aligned}
    \mathcal{M}^{\text{open}}_R &= T^{c_3}_{i_1i_2} \overline{v}_{s_1}(\mathbf{p}_1) \left\{\mathbb{P}_R \left[ \tau_{1,\mu} A_{1,m_W}^{\textup{open}} + \tau_{2,\mu} A_{2,m_W}^{\textup{open}} \right]  
		+\right.\\&\left.
		+ \frac{2}{\cos^4 \theta_W}\left( T_q - Q_q \sin^2 \theta_W \right)^2 \mathbb{P}_R \left[ \tau_{1,\mu} A_{1,m_Z}^{\textup{open}} + \tau_{2,\mu} A_{2,m_Z}^{\textup{open}} \right]\right\} u_{s_2}(\mathbf{p}_2) \epsilon^{\lambda_3}_\mu(\mathbf{p}_3)  \,,
	\end{aligned}
\end{align}
\begin{align}
    \begin{aligned}
	\mathcal{M}^{\text{closed}} & = T^{c_3}_{i_1i_2} \frac{1}{2} \overline{v}_{s_1}(\mathbf{p}_1) \left\{ 4 \left[ \tau_{1,\mu} F_{1,m_W}^{\textup{closed}} + \tau_{2,\mu} F_{2,m_W}^{\textup{closed}} \right] + \frac{2}{\cos^4 \theta_W}  \times\right.\\&\left. 
	\times\left( \frac{5}{4}-\frac{7}{3}\sin^2 \theta_W+\frac{22}{9}\sin^4 \theta_W \right) \left[ \tau_{1,\mu} F_{1,m_Z}^{\textup{closed}} + \tau_{2,\mu} F_{2,m_Z}^{\textup{closed}} \right] \right\} u_{s_2}(\mathbf{p}_2) \epsilon^{\lambda_3}_\mu(\mathbf{p}_3)  \,,
	\end{aligned}
\end{align}
\begin{align}
    \tau_{1,\mu} = \slashed{p}_3 p_{2\mu} - p_2 \cdot p_3 \gamma_\mu  \,,\qquad\qquad\tau_{2,\mu} = \slashed{p}_3 p_{1\mu} - p_1 \cdot p_3 \gamma_\mu \,,
\end{align}
for the quark-initiated process. Notice that in the case with an open quark line the rescaling of the coupling constants depends explicitly on the quarks polarization.

Once the tensor structures of the amplitudes have been determined we construct and apply projectors to extract the form factors $F_j$ and $A_j$.

\section{Evaluation of the form factors}
\label{MIs}

\begin{figure}
\centering
	\subfloat[T1: $\FINB{1,1,1,1,0,1,1,1,0}$]{{\includegraphics[width=0.30\textwidth]{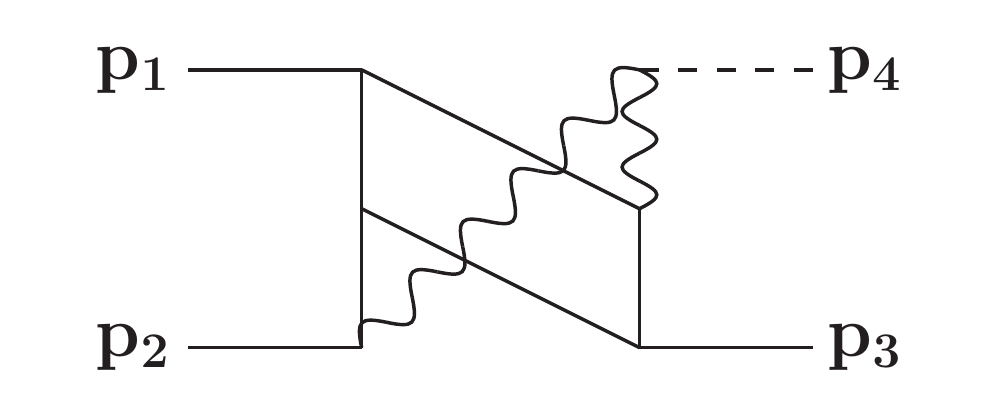}}\label{fig:T1}}
	\qquad
	\subfloat[T2: $\FINA{1,1,1,1,1,1,1,0,0}$]{{\includegraphics[width=0.30\textwidth]{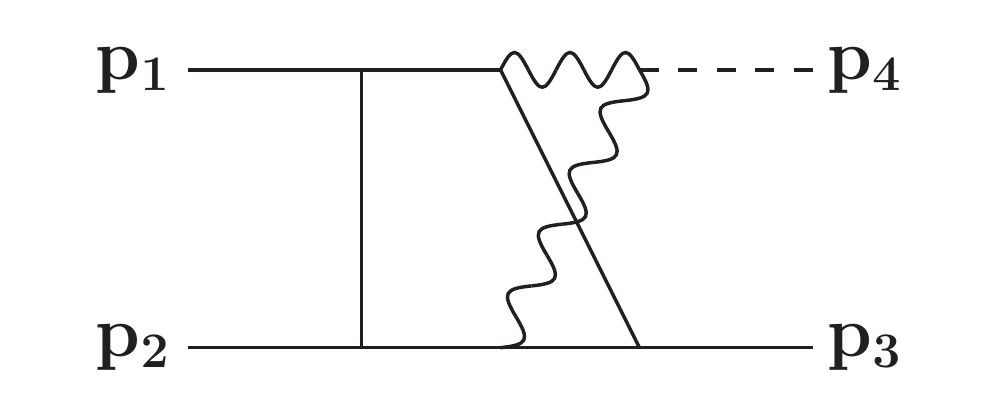}}\label{fig:T2}}
	\\
	\subfloat[T3: $\FIPL{0,1,1,1,1,0,1,1,1}$]{{\includegraphics[width=0.30\textwidth]{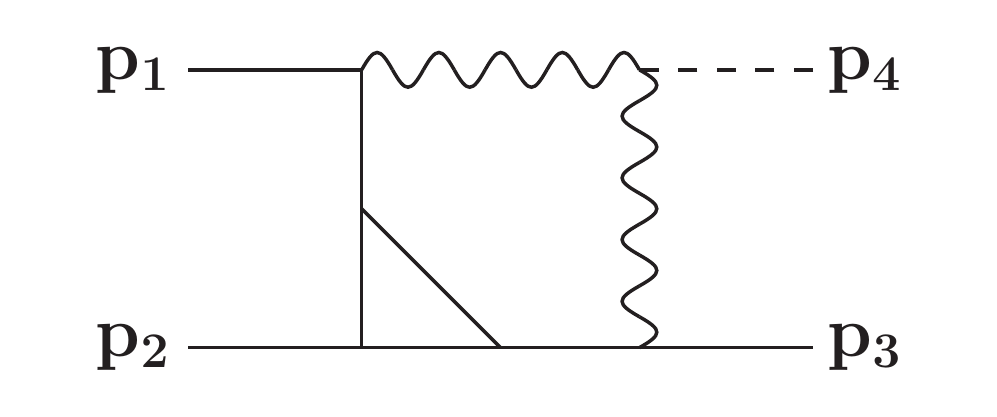}}\label{fig:T3}}
	\qquad
	\subfloat[T4: $\FIPL{1,1,1,1,0,0,1,1,1}$]{{\includegraphics[width=0.30\textwidth]{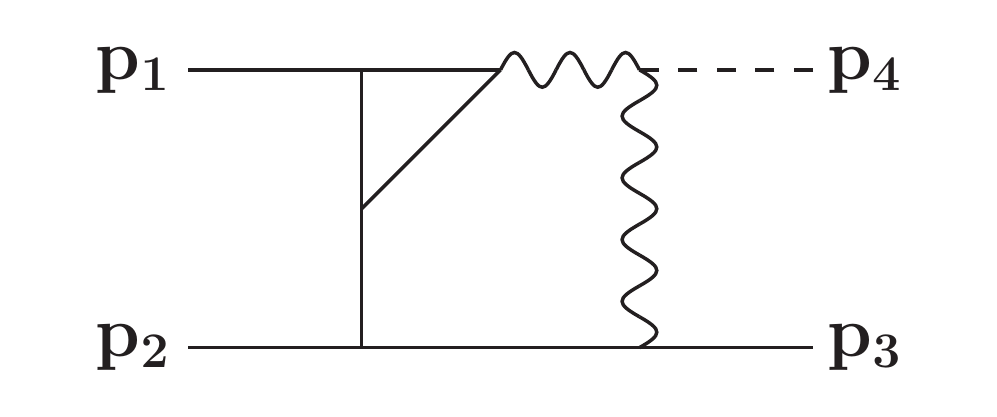}}\label{fig:T4}}
	\\
	\subfloat[T5: $\FIPL{0,1,1,1,1,1,1,1,0}$]{{\includegraphics[width=0.30\textwidth]{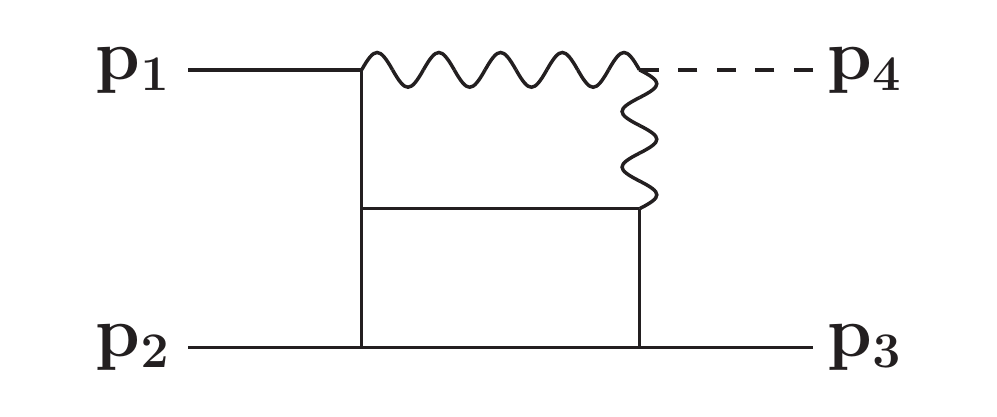}}\label{fig:T5}}
	\qquad
	\subfloat[T6: $\FIPL{1,1,1,1,1,1,1,0,0}$]{{\includegraphics[width=0.30\textwidth]{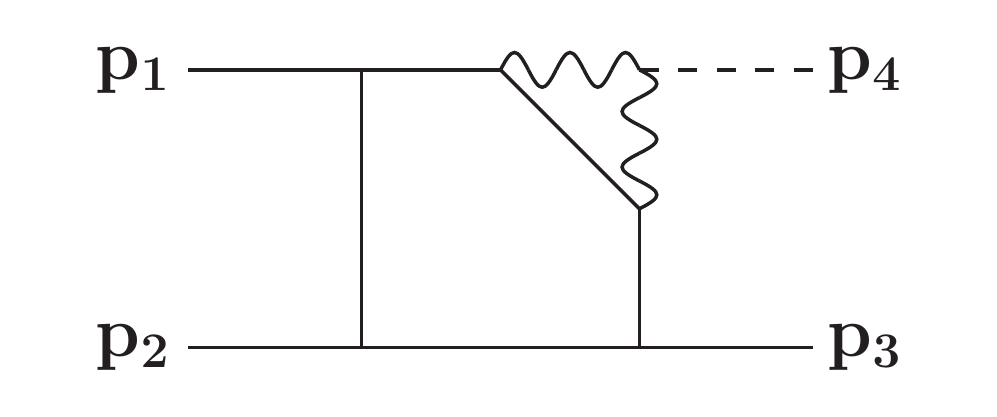}}\label{fig:T6}}	
	\caption{The six top sectors appearing in the amplitude. Straight (wavy) lines denote massless (massive) propagators. The dashed line indicates the Higgs boson.}
    \label{fig:topolo}
\end{figure}

The form factors $F_j$ and $A_j$ are extracted by applying projectors to the amplitudes, and are expressed in terms of two-loop Feynman integrals, whose top sectors, up to permutations of the external legs, are depicted in Figure~\ref{fig:topolo}. The $ggHg$ channel can be described in terms of the T2 and T6 top sectors only, why the $q\overline{q}Hg$ contribution requires all the six top sectors. As a consequence, the $ggHg$ amplitude consists of 61 master integrals, while the $q\overline{q}Hg$ amplitude requires 30 more. The presence of internal massive lines associated to the Electroweak vector bosons generates the eight different square roots
\begin{align}
    \begin{aligned}
        r       &= m_h^2 \sqrt{1-4 m_V^2/m_h^2} \,,\qquad&&
        r_{ust} &= \sqrt{s^2 u^2 +2su(t-s)m_V^2+(s+t)^2m_V^4}    \,,\\
        r_t     &= \sqrt{r^2-4 m_V^2 su/t}      \,,\qquad&&
        r_{sut} &= \sqrt{s^2 u^2+2su(t-u)m_V^2+(t+u)^2m_V^4}     \,,
    \end{aligned}
\end{align}
\begin{align}
    \begin{aligned}
        r_u     &= \sqrt{r^2-4 m_V^2 st/u}      \,,\qquad&&
        r_{stu} &= \sqrt{s^2 t^2 + 2st(u-t)m_V^2+(t+u)^2m_V^4}   \,,\\
        r_{tu}  &= \sqrt{1-4 m_V^2 /(t+u)}      \,,\qquad&&
        r_{uts} &= \sqrt{u^2 t^2 + 2ut(s-t)m_V^2+(s+t)^2m_V^4}   \,,
    \end{aligned}
\end{align}
which, to our knowledge, are not all simultaneously rationalisable. We observe though, that for a single master integral no more than three square roots appear, and in all these cases these sub-sets can be simultaneously rationalised.

We derive the differential equations w.r.t.\ $t$, $u$, $m_h^2$, and $m_V^2$ for the master integrals and we manage to obtain an $\epsilon$-factorized $\td\log$-form for the matrix of coefficients in the $ggHg$ case. Despite this encouraging result, to evaluate the top-sector master integrals we need to combine expressions containing different parametrizations, ultimately leading to very large results which made the computation of the highest $\epsilon$-orders practically impossible.\footnote{It is still possible to solve the system of differential equations in terms of GPLs without square roots when considering the planar top sector T6 only, as done in \cite{Becchetti:2018xsk}.} We turn then to the direct integration of the master integrals over Feynman--Schwinger parameters. This approach saves us from integrating over square roots thanks to the fact that our top sectors (and consequently the master integrals) are \emph{linearly reducible} \cite{Panzer:2014caa}, i.e.\ there exists an integration order over the Feynman--Schwinger parameters such that each integration returns a hyperlogarithm of rational argument in the next integration variable. In this way we always integrate over $\td\log$ kernels, obtaining results expressed in terms of  multiple polylogarithms (also known as Goncharov polylogarithms, or GPLs), and integration over square roots might appear at most during the last integration, dispensing us from rationalising the expressions.\footnote{The GPLs will contain explicit square roots in the arguments, but this poses no issues.}

\begin{figure}
\centering
	\begin{align*}
		\imineq{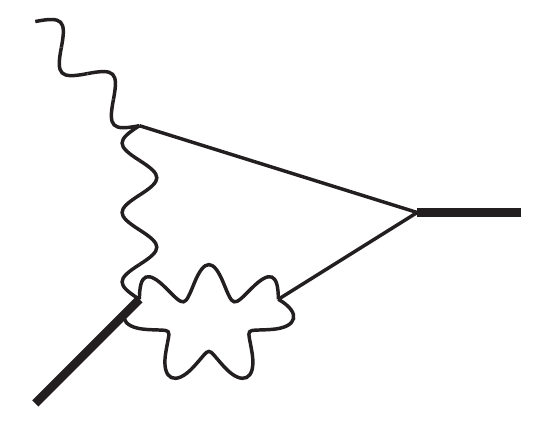}{10}
		\,\,\Rightarrow\,\,
		\imineq{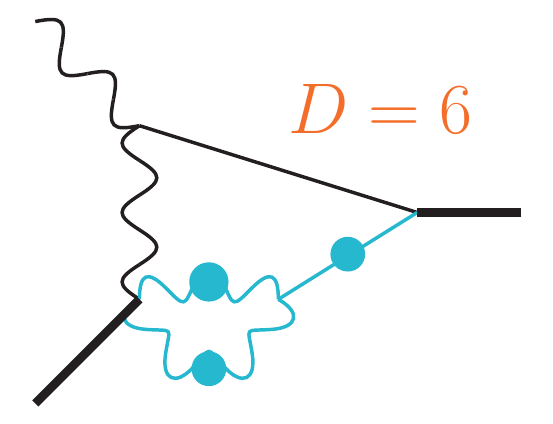}{10} + \imineq{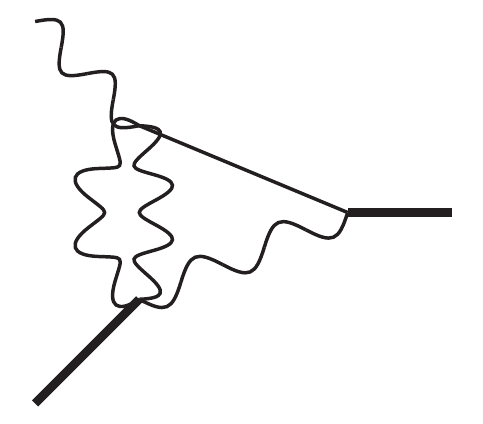}{8} + \imineq{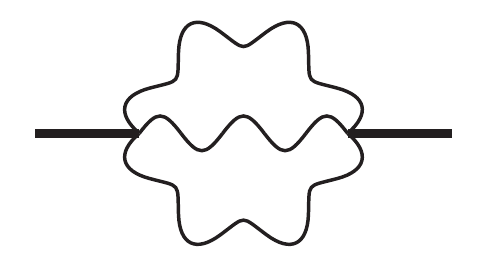}{4} + \imineq{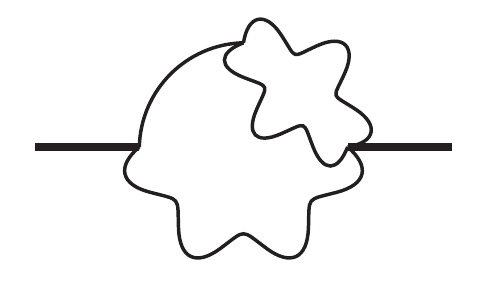}{4} + \dots
	\end{align*}	
	\caption{Decomposition of the master integral $\FIPL{0,1,1,0,0,1,1,1,0}$ in terms of a finite candidate and (divergent) sub-graphs.}
    \label{fig:tofb}
\end{figure}

In full generality, our two-loop master integrals can present divergencies in the dimensional regularisation parameter $\epsilon = (4-d)/2$ up to $\epsilon^{-4}$, while we expect the $ggHg$ amplitude to be finite and the $q\overline{q}Hg$ one to show at most $\epsilon^{-2}$ poles. In order to exploit the milder pole structure of the amplitudes we look for a (quasi-)finite basis of master integrals, constructing candidates by removing both UV and IR divergencies: UV divergencies can be removed by raising the power of massive denominators until a negative superficial degree of divergence is reached, while IR divergencies are mitigated by shifting dimensions from $d=4-2\epsilon$ to $d=6-2\epsilon$ \cite{Tarasov:1996br,Lee:2009dh} (cfr. Fig.~\ref{fig:tofb}). The removal of IR divergencies can be understood considering that the dimensional shift generates a Gram determinant, which is essentially a collection of scalar products containing loop momenta cancelling the IR singularities coming from the scalar products in the denominators. By applying this procedure to the master integrals with more than four propagators we manage to decompose them into a finite integral plus some divergent sub-graphs. This decomposition can produce explicit pole cancellation already before substituting the expressions of the master integrals, and in general allows for fewer $\epsilon$-orders of the master integrals to be computed. As a last remark, attention has to be taken in the choice of a (quasi-)finite basis that does not worsens the $\epsilon$-pole structure of the coefficients of the master integrals themselves, ultimately resulting in more orders of to be computed.

We compute the analytic $\epsilon$-expansions of the master integrals using {\HyperInt} \cite{Panzer:2014caa} and insert them in the form factors. In the $q\overline{q}Hg$ case we proceed by renormalizing $\alpha_S$ in the $\overline{\text{MS}}$ scheme and removing the IR poles by subtracting the LO amplitude multiplied by the Catani operator $\mathrm{I}^{(1)}_{q\overline{q}g}$, which describes the universal IR behavior of QCD, see \cite{Catani:1996vz}.

The $\epsilon$-finite parts of the form factors are still large expressions difficult to handle at this stage. Such results are linear combinations of GPLs, whose coefficients are algebraic expressions of the kinematics and the masses. The first step towards the simplification of these expressions is to find a basis of algebraic prefactors. This procedures starts applying the partial fraction decomposition algorithm described in \cite{Heller:2021qkz} to find a basis of algebraic "monomials" without inserting spurious poles, followed by the reduction of the prefactors to a basis using the common set of monomials as a vector space. Now that no further relations can be found among the prefactors, we numerically search for zeroes and null linear combinations among the transcendental expressions, using the results also to express higher weight GPLs in terms of lower weight ones. The expressions obtained after this process are smaller, both in size and number of functions, shrinking from more than $1\,\text{GiB}$ to about $1\,\text{MiB}$, see Table~\ref{table:simpli} for details.

\begin{table}
	\centering
    \small
	\begin{tabular}{cccccc}
\toprule
~       &Original	&{Partial reduction} &{Monomial reduction}  &{Basis}		&{No zeroes}	\\
\midrule
$\tilde{F}_{1,\epsilon^0}^{\textup{open},(1)}$ &8~(31)  &7~(31)  &7~(24)  &6~(24)  &6~(24)  \\
$\tilde{F}_{2,\epsilon^0}^{\textup{open},(1)}$ &8~(33)  &7~(33)  &7~(30)  &6~(30)  &5~(24)  \\
$\tilde{F}_{1,\epsilon^1}^{\textup{open},(1)}$ &23~(70) &18~(70) &18~(50) &12~(50) &9~(21)	 \\
$\tilde{F}_{2,\epsilon^1}^{\textup{open},(1)}$ &22~(58) &16~(58) &16~(43) &12~(43) &9~(22)	 \\
$\tilde{F}_{1,\epsilon^2}^{\textup{open},(1)}$ &45~(87) &26~(87) &26~(65) &17~(65) &12~(37) \\
$\tilde{F}_{2,\epsilon^2}^{\textup{open},(1)}$ &44~(73) &23~(73) &23~(58) &17~(58) &10~(22) \\
\midrule
$\tilde{E}_{1}$ &46~(45)	&22~(45) &22~(28) &15~(28) &10~(17)	     \\
$\tilde{E}_{2}$ &46~(45) &22~(45) &22~(28) &15~(28) &10~(17)      \\
$\tilde{F}_{1,\epsilon^0}^{N_c}$		     &1410~(1454) &234~(1294)  &234~(1093)  &134~(1093) &100~(983)  \\
$\tilde{F}_{2,\epsilon^0}^{N_c}$		     &1413~(1389) &213~(1285)  &213~(1186)  &134~(1186) &117~(1169) \\
$\tilde{F}_{1,\epsilon^0}^{1/N_c}$	         &5526~(6789) &1174~(6788) &1100~(5177) &690~(3823) &325~(983)  \\
$\tilde{F}_{2,\epsilon^0}^{1/N_c}$	         &5524~(5905) &1139~(5894) &1139~(4604) &784~(4517) &460~(1169) \\
\midrule
$\tilde{B}_{1}$	  &4~(7)    &4~(7)	  &4~(7)    &4~(7)	  &~4(7)    \\
$\tilde{B}_{2}$	  &4~(7)	&4~(7)    &4~(7)    &4~(7)	  &~4(7)    \\
$\tilde{C}_{1}$   &43~(104) &35~(104) &32~(98)	&30~(98)  &30~(95)  \\
$\tilde{C}_{2}$   &41~(188) &34~(188) &33~(184)	&30~(184) &30~(123) \\
$\tilde{D}_{1}$   &67~(161) &64~(161) &61~(151) &54~(151) &54~(151)      \\
$\tilde{D}_{2}$   &93~(601) &91~(601) &89~(513) &54~(346) &54~(136)      \\

\bottomrule
	\end{tabular}
	\caption{Number of linearly-independent rational prefactors (and rational monomials) at different stages of the reduction procedure for the $q\overline{q}Hg$ amplitude. We list the first three non-zero orders $\tilde{A}$ in the $\epsilon$ expansion for the LO amplitude, followed by the different component of the two-loop NLO amplitude and then by the different parts of the two-loop finite remainder.}
	\label{table:simpli}
\end{table}

As a last step for the $ggHg$ case we use symbol techniques to rewrite the GPLs up to weight 3 in terms of $\log$s, $\Li{2}$s, and $\Li{3}$s, and make their analytic continuation explicit in the physical region $s>m_h^2>0$, $t,u<0$, $m_V^2<m_h^2<4m_V^2$. In this way we obtain expressions optimized for fast and stable numerical evaluations.

To provide physical results we construct the helicity amplitudes for the $ggHg$ and $q\overline{q}Hg$ processes. We prefer such expressions over the form factors since they are more directly related to observables and so present in general a simpler structure.

In the $ggHg$ case, only two non-zero independent helicity amplitudes exist:
\begin{align}
    \mathcal{A}^{ggHg}_{+++} &= \frac{m_h^2}{\sqrt{2}\langle12\rangle\langle23\rangle\langle31\rangle} \frac{s u}{m_h^2} \left( \mathcal{F}_1 + \frac{t}{u}\mathcal{F}_2 + \frac{t}{s}\mathcal{F}_3 + \frac{t}{2}\mathcal{F}_4 \right) ,
    \\
    \mathcal{A}^{ggHg}_{++-} &= \frac{[12]^3}{\sqrt{2}m_h^2[13][23]} \frac{u m_h^2}{s} \left( \mathcal{F}_1 + \frac{t}{2}\mathcal{F}_4 \right) .
\end{align}
For the $q\overline{q}Hg$ we have also have only two non-zero independent helicities:
\begin{align}
    \begin{aligned}
        \mathcal{A}^{q\overline{q}Hg}_{RL+} &= \left\{\frac{1}{2}\left[ 4 \mathcal{F}_{1,m_W}^{\textup{closed}} 
	+ \frac{2}{\cos^4 \theta_W}\left( \frac{5}{4}-\frac{7}{3}\sin^2 \theta_W+\frac{22}{9}\sin^4 \theta_W \right) \mathcal{F}_{1,m_Z}^{\textup{closed}} \right]
			+\right.\\&\qquad\left.+
			\left[ \mathcal{F}_{1,m_W}^{\textup{open}} + \frac{2}{\cos^4 \theta_W}\left( T_q - Q_q \sin^2 \theta_W \right)^2 \mathcal{F}_{1,m_Z}^{\textup{open}} \right] \right\} \frac{s}{\sqrt{2}}\frac{[23]^2}{[12]} ,
	\end{aligned}
\end{align}
\begin{align}
    \begin{aligned}
    \mathcal{A}^{q\overline{q}Hg}_{LR+} &= \left\{\frac{1}{2}\left[ 4 \mathcal{F}_{2,m_W}^{\textup{closed}} 
	+ \frac{2}{\cos^4 \theta_W}\left( \frac{5}{4}-\frac{7}{3}\sin^2 \theta_W+\frac{22}{9}\sin^4 \theta_W \right) \mathcal{F}_{2,m_Z}^{\textup{closed}} \right]
			+\right.\\&\qquad\left.+
			\left[ \frac{2}{\cos^4 \theta_W} Q_q^2 \sin^4 \theta_W \mathcal{F}_{2,m_Z}^{\textup{open}} \right]  \right\} \frac{s}{\sqrt{2}}\frac{[13]^2}{[12]} .
	\end{aligned}
\end{align}
We observe a weight drop in $\mathcal{A}^{ggHg}_{+++}$ with respect to the other helicity amplitudes: only GPLs up to weight 3 are present here, instead of weight 4. This agrees with what was observed at two and three loops in the $ggH$ case, where terms only up to weight 3 and 5, respectively, were observed \cite{Bonetti:2016brm,Bonetti:2017ovy}. Another important observation for the $q\overline{q}Hg$ case is that for symmetry reasons we expect $\mathcal{F}_{1,m_V}^{\text{open/closed}}$ to be equal to $\mathcal{F}_{2,m_V}^{\text{open/closed}}$ under a $t \leftrightarrow u$ exchange. We do not impose of this property explicitly, rather we use it to cross-check our results along the simplification chain described above.

\section{Conclusions \& outlook}
\label{conout}

The analytic calculations described in these proceedings are the last missing ingredient of the NLO light-quarks mixed QCD-Electroweak corrections to Higgs boson production at the LHC. This paves the way to the computation of the NLO hadronic cross section including Electoweak effects, completing what was started in \cite{Becchetti:2021axs}, where only the gluon-initiated case was considered at NLO. A complementary study will focus instead on the inclusion of top-quark contributions, in order to fully address the effects and issues related to massive fermions coupling to Electroweak bosons at NLO. This next step poses new and non-trivial challenges, such as non-factorisable terms proportional to $\gamma_5$ and complex integral reductions and master integrals evaluation, which will now depend on five parameters and contain many massive lines.

~

\textbf{Acknowledgments.}~These contributions are based on \cite{Bonetti:2020hqh,Bonetti:2022lrk}, by Erik Panzer, Vladimir A.\ Smirnov, and Lorenzo Tancredi. M.B.\ is supported by the Deutsche Forschungsgemeinschaft (DFG) under grant no.\ 396021762 - TRR 257.

\bibliographystyle{JHEP}
\bibliography{Biblio}

\end{document}